\documentclass[sigconf,authorversion]{acmart}

\newif\ifauthor
\authortrue

\usepackage[utf8]{inputenc}
\usepackage[english]{babel}

\usepackage{graphicx}
\graphicspath{{figures/}} 
\DeclareGraphicsExtensions{.pdf,.jpg,.jpeg,.png,.JPG}

\usepackage{amsmath}
\usepackage[printwatermark]{xwatermark}
\usepackage{balance}
\usepackage{url}
\usepackage{bytefield}
\usepackage{booktabs}
\usepackage{multirow}
\usepackage{xspace}
\usepackage{wrapfig}           
\usepackage[export]{adjustbox} 

\addto\extrasenglish{}
\addto\extrasenglish{}
\addto\extrasenglish{}
\addto\extrasenglish{}
\addto\extrasenglish{}

\newcommand{\riscy}{RI5CY\xspace}
\newcommand{\riscv}{\mbox{RISC-V}\xspace}

\begin{document}


\copyrightyear{2018}
\acmYear{2018}
\setcopyright{acmlicensed}
\acmConference[ACSAC '18]{2018 Annual Computer Security Applications Conference}{December 3--7, 2018}{San Juan, PR, USA}
\acmBooktitle{2018 Annual Computer Security Applications Conference (ACSAC '18), December 3--7, 2018, San Juan, PR, USA}
\acmPrice{15.00}
\acmDOI{10.1145/3274694.3274728}
\acmISBN{978-1-4503-6569-7/18/12}

\title{Pointing in the Right Direction - Securing Memory Accesses in a Faulty World}

\author{Robert Schilling}
\affiliation{
	\institution{Graz University of Technology}
	\institution{Know-Center GmbH}
}
\email{robert.schilling@iaik.tugraz.at}

\author{Mario Werner}
\affiliation{\institution{Graz University of Technology}}
\email{mario.werner@iaik.tugraz.at}

\author{Pascal Nasahl}
\affiliation{\institution{Graz University of Technology}}
\email{pascal.nasahl@student.tugraz.at}

\author{Stefan Mangard}
\affiliation{\institution{Graz University of Technology}}
\email{stefan.mangard@iaik.tugraz.at}

\begin{abstract}
Reading and writing memory are, besides computation, the most common operations
a processor performs. The correctness of these operations is therefore essential
for the proper execution of any program. However, as soon as fault attacks are
considered, assuming that the hardware performs its memory operations as
instructed is not valid anymore. In particular, attackers may induce faults with
the goal of reading or writing incorrectly addressed memory, which can have
various critical safety and security implications.

In this work, we present a solution to this problem and propose a new method for
protecting every memory access inside a program against address tampering. The
countermeasure comprises two building blocks. First, every pointer inside the
program is redundantly encoded using a multi-residue error detection code. The
redundancy information is stored in the unused upper bits of the pointer with
zero overhead in terms of storage. Second, load and store instructions are
extended to link data with the corresponding encoded address from the pointer.
Wrong memory accesses subsequently infect the data value allowing the software
to detect the error.

For evaluation purposes, we implemented our countermeasure into a RISC-V
processor, tested it on a FPGA development board, and evaluated the induced
overhead. Furthermore, a LLVM-based C compiler has been modified to
automatically encode all data pointers, to perform encoded pointer arithmetic,
and to emit the extended load/store instructions with linking support. Our
evaluations show that the countermeasure induces an average overhead of 10\,\%
in terms of code size and 7\,\% regarding runtime, which makes it suitable for
practical adoption.
\end{abstract}


\maketitle

\section{Introduction}

A memory access is a highly critical operation. Many decisions inside a program
rely on the correct execution of a memory access. Password checks, signature
verification, or grants to a privileged function they all rely on the genuine
execution of a memory access.

Under normal operating conditions, a memory access reads/writes from/to the
desired location and random malfunctions, e.g., caused by cosmic
radiation~\cite{baumann2005radiation}, are comparably rare. However, the
situation changes dramatically as soon as intentionally induced faults, via
so-called fault attacks, are considered. Here, the attacker modifies the state
of a computing device by, e.g., inducing glitches on the voltage supply or the
clock signal~\cite{DBLP:journals/pieee/Bar-ElCNTW06} or by shooting with a laser
on the chip~\cite{DBLP:conf/cardis/SelmkeBHS15}. Such a fault attack is capable
of skipping instructions~\cite{DBLP:conf/ccs/BreierJC15}, redirecting the memory
access~\cite{derouet2007secure}, or flipping bits in registers or memory leading
to a critical attack vector~\cite{DBLP:conf/cardis/GiraudT04}. While this type
of attack requires local access to the device to induce a fault, more advanced
attacks can even induce faults remotely. For example, the Rowhammer
effect~\cite{DBLP:conf/isca/KimDKFLLWLM14}, which modifies the state of the
memory by frequently accessing neighboring memory cells, can also be induced in
software via Javascript~\cite{DBLP:conf/dimva/GrussMM16} or remotely over a
network interface~\cite{tatar2018throwhammer,lipp2018nethammer}.

While a fault may not directly reveal sensitive information, different
techniques have been developed to exploit faulty computation. For example, it
has been shown that it is possible to deduce the secret key in various
cryptographic algorithms~\cite{DBLP:journals/joc/BonehDL01,
DBLP:journals/jce/AliMT13} solely by analyzing the faulty computation output.
Subsequently, a lot of research has been performed to protect specific
cryptographic algorithms against fault attacks
~\cite{DBLP:conf/cases/BarenghiBKPR10,DBLP:conf/fdtc/RauzyG14}. However, the
hardening of general purpose software against fault attacks is a young research
area.

Two complementary subareas exist. The first subarea deals with the protection
of the executed code. The respective
techniques~\cite{DBLP:journals/compsec/ClercqGUMV17,
DBLP:journals/corr/abs-1802-06691, DBLP:journals/corr/abs-1803-08359} typically
enforce control-flow integrity~(CFI), which is also a well-known mitigation
strategy against software attacks, with fine granularity. The resulting
countermeasures ensure that executed instructions and branches are genuine and
that they are processed in the correct sequence without omission.

The second subarea mainly deals with the protection of data. There, well-known
redundancy-based techniques like arithmetic
codes~\cite{DBLP:journals/tc/Brown60,massey1964survey,DBLP:journals/tit/RaoG71,
DBLP:conf/safecomp/SchiffelSSF10} are utilized. In these schemes, the data is
encoded into a redundant domain, where faults are detectable up to a certain
number of bit flips. Interestingly, while such schemes were initially developed
to protect the data while it is stored in the memory, arithmetic codes also
support to perform certain arithmetic operations on the encoded value.


However, even when mechanisms of the two subareas are combined, i.e., a system
implements a CFI protection mechanism and redundantly encodes the data, memory
transfers from the processor via the memory subsystem are still vulnerable to
fault attacks. Namely, when a fault modifies the address on one of the buses
during the read or write operation, the data is read or written from/to the
wrong memory location, which is not trivially detectable given that the data is
unmodified. Similar effects can be triggered by injecting faults into pointers,
which are typically not prevented by these schemes.

Unfortunately, current extensions to data encoding, which aim to solve this
issue, are very costly and impose severe restrictions on the protected code.
ANB-codes~\cite{DBLP:conf/safecomp/SchiffelSSF10}, for example, introduce a
tremendous runtime overhead of 90\,\% on average on top of already expensive
AN-codes, solely to solve the memory access problem. Furthermore, they can only
protect a limited set of variables with well-known memory alignment and size.
More efficient and less restrictive approaches are needed to protect memory
accesses against address tampering.

\subsection*{Contribution}

In this work, we address the issue of unprotected memory accesses in the context
of fault attacks. We propose a practical solution to detect address tampering in
pointers and on memory buses. Our generic approach works independently of the
used code and data protection schemes and therefore can effectively be combined
with state-of-the-art techniques in the context of hardening general purpose
computing against fault attacks.

In detail, the contributions of this paper are as follows. First, we present a
new approach to protect pointers against faults with negligible overhead in
terms of runtime and storage requirements. We encode pointers using a
multi-residue arithmetic code, which allows us to detect faults on encoded
pointers during both storage and computation. The redundancy information of the
code word is hereby stored in the unused upper bits of a pointer to fully
utilize the available register space and yield zero-overhead for storing an
encoded pointer. Furthermore, by transforming the pointer arithmetic into the
encoded multi-residue domain, the protection of the pointer is maintained also
when performing arithmetic operations on the pointer; e.g., when adding an
offset to the stack pointer.

Second, we propose an efficient way to protect memory accesses from tampering by
linking the stored data in memory with the address of the access. We establish
this link whenever data is written to the memory and remove the link as soon as
the data is read back into the processor. When considering fault attacks,
countermeasures like data encoding are already necessarily employed. By linking
the redundant address information with the encoded data, faults during
addressing manifest as errors in the redundantly encoded data, where they can be
detected. As the result, data integrity checks implicitly also checks for
address tampering and make explicit addressing error checks unnecessary.

Finally, to evaluate the concept, we integrated our protection mechanism into an
FPGA hardware implementation of an open-source \riscv processor. Furthermore, to
avoid tedious manual encoding of all pointers and addresses inside the program,
we integrated this concept directly into a LLVM-based C compiler, which is
capable of automatically protecting complex codebases without manual
interference. The resulting prototype induces 10\,\% code size and less than
7\,\% runtime overhead on average.

\subsection*{Outline}

The remainder of this paper is structured as follows. \autoref{sec:Background}
discusses the threat model and the attack vector, gives an introduction to
arithmetic codes, and discusses related work. In \autoref{sec:pointer}, we
describe how we protect pointers against fault attacks. The approach to link the
pointer protection with data encoding is presented in~\autoref{sec:memory}.
\autoref{sec:architecture} details how we extend the \riscv instruction set to
support encoded pointers and discusses our compiler modifications. Finally,
\autoref{sec:evaluation} evaluates the overhead and \autoref{sec:conclusion}
concludes this work.

\section{State of the Art and Background}
\label{sec:Background}

In this section, we first describe the attack vector and the threat model we
consider. Furthermore, we present state-of-the-art methods of error detection
codes, which we use to protect a memory access against tampering efficiently. We
also show related concepts, which aim to secure pointers or a memory access in
general.

\subsection{Threat Model and Attack Vector}
\label{sec:fault_model}

For this work, we assume a powerful attacker, which performs fault attacks in
order to compromise a system. Faults can be induced into instructions and data
at various places like, for example, in registers, during computation in the
ALU, on buses, and in memory. Many of these attack vectors can be covered by
existing and established countermeasures, which we assume to be in place.
Namely, CFI-based fault
countermeasures~\cite{DBLP:journals/compsec/ClercqGUMV17,
DBLP:journals/corr/abs-1802-06691, DBLP:journals/corr/abs-1803-08359}, which
enforce the authenticity of instructions as well as their execution sequence,
can be used to protect code against faults. Furthermore, such a CFI scheme
already protects function pointers, which do not require further protection.
Data, on the other hand, can be protected during computation and storage using
data encoding techniques like, for example, arithmetic
codes~\cite{DBLP:journals/tc/Brown60,massey1964survey,DBLP:journals/tit/RaoG71,
DBLP:conf/safecomp/SchiffelSSF10}. However, as soon as data is transferred via
a memory bus these codes are insufficient. Namely, while the value itself is
protected via the code, the corresponding address information remains
vulnerable. Furthermore, pointers as such, typically, also remain unprotected by
the data encoding schemes considering that eventually the plain value of the
pointer is used to address the memory.

\begin{figure}[t]
	\centering
	\includegraphics[width=0.8\columnwidth]{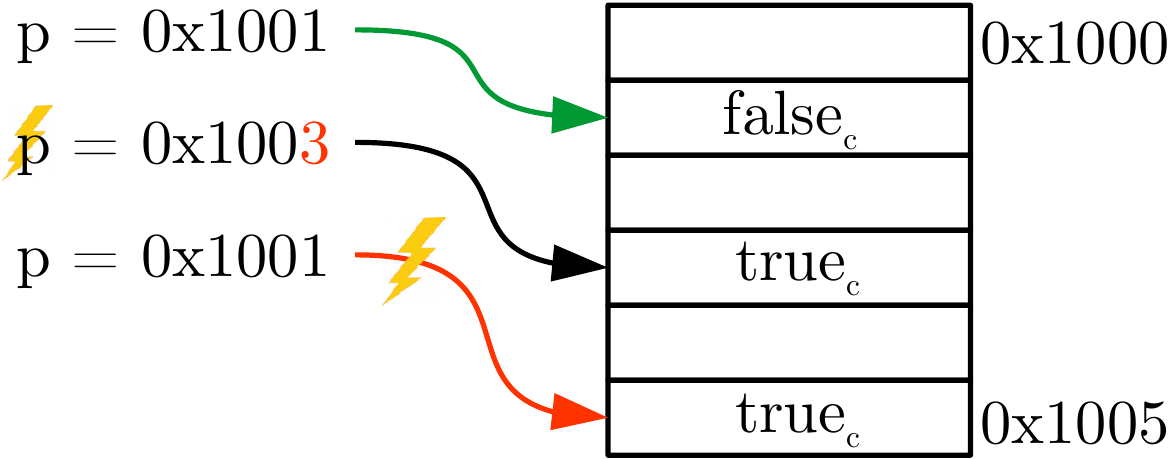}
	\caption{Attack vector: Modified pointers and manipulated memory accesses.}
	\label{fig:attack_vector}
\end{figure}

To illustrate the problem, \autoref{fig:attack_vector} visualizes a simple
memory access. On the left side there is the pointer used for a memory access,
on the right side there is the memory, and the arrow in between denotes the
memory access. The data in the memory is redundantly encoded denoted by the
\textit{c}-subscript of the variables. Originally, the pointer $p$ points to the
address 0x1001 to readout the value false\textsubscript{c} from the memory.
However, a fault can manipulate the memory access to readout a wrong value. In
particular, there are two error sources, which can lead to a wrong memory
access. First, the attacker can modify the pointer as shown in the middle
example in \autoref{fig:attack_vector}. If a pointer gets modified, then all
subsequent memory accesses lead to a wrong location. An attacker could, e.g.,
modify two pointers used for a signature comparison to point to the same
location, which always bypasses the memory comparison. This can occur anywhere
in the program, also during pointer arithmetic. The second source of a
manipulated memory access is the memory operation itself. When assuming the
pointer is correct and not manipulated, the memory access can still be
manipulated. A fault on the address bus can redirect the memory access to a
wrong location as indicated in the third example.

Both of these attack vectors can lead to a wrong memory access. Today, there is
no efficient way to protect them leaving frequently used memory operations
completely unprotected against fault attacks.

\subsection{Error Detection Codes}
\label{sec:codes}

Error detection codes~\cite{peterson1961codes} are a well-known and well-studied
concept to detect errors during storage or transmission. However, different
types of code have been developed, which also support the computation on encoded
data. Logical operations, for example, can directly be computed in the encoded
domain when binary linear codes~\cite{hamming1950error} are used. Arithmetic
codes, on the other hand, can be used when primarily arithmetic operations have
to be performed on encoded data.

\subsubsection{AN(B)-Codes.}

AN-codes~\cite{DBLP:journals/tc/Brown60,forin1990vital} are an example for such
an arithmetic code and
are defined by multiplying the functional value $x$ with the encoding constant
$A$: $x_c = x \cdot A$. Therefore, all code words are multiples of the encoding
constant $A$ and every value in between corresponds to an invalid code word. To
check if a code word is valid, a modulo operation with the encoding constant is
performed, which must yield zero. Decoding is done by using an integer division
with the encoding constant. Because of multiplying the functional value with the
encoding constant, it cannot be separated from the redundancy part, thus the
name \textit{non-separable} code. The encoding constant $A$ defines the error
detection capabilities. Finding a good encoding constant is not easy and
currently only possible via exhaustive search~\cite{DBLP:conf/wisa/MedwedS09}.
Research already found suitable encoding constants, which maximize the error
detection capabilities, so-called \textit{Super
$A$}~\cite{DBLP:conf/hase/HoffmannUDSLS14}. To maintain the error detection
capabilities, the functional value needs to be less than the encoding constant,
which limits the general-purpose application of this code in real-world
applications. Furthermore, this type of code does not protect the address in a
memory access.

Forin and Schiffel et al.~\cite{forin1990vital,DBLP:conf/safecomp/SchiffelSSF10}
extend AN-codes by assigning a variable dependent signature $B_x$ to each
encoded variable $x_c$. This yields the encoding formula $x_c = A \cdot x + B_x$
with $B_x < A$. By adding the variable dependent signature $B_x$ to the AN-code,
the AN-code property that all encoded values are a multiple of $A$ is
intentionally destroyed. Since $B_x$ is less than $A$, decoding works the same
as for normal AN-codes using an integer division. A check is also performed
using a modulo operation with the encoding constant, which now must yield the
signature $B_x$. A compiler keeps track of all assigned signatures and is able
to insert checks for the modified ANB code words. By assigning a
variable-dependent signature to the code words, a wrong memory access can be
detected, as long as signatures do not cancel out due to arithmetic. However,
this approach turns out to be complicated in practice. On the one hand,
ANB-codes have a performance penalty of 90\,\% on average on top of AN-codes. On
the other hand, the value of the signature $B_x$ needs to be less than the
encoding constant which limits the number of variables to be encoded.

\subsubsection{Residue Codes.}

A different class of arithmetic codes are residue codes~\cite{massey1964survey}.
Here, a residue code word $x_c$ is defined by the tuple $x_c = (x, r_x = M |
x)$, where $x$ denotes the functional value and $r_x$ the residue. The residue
$r_x$ is hereby computed as the remainder $M | x$ with respect to the code
modulus $M$. Residue codes separate the redundancy part from the functional
value $x$ and therefore are called \textit{separable} codes. Although the
modulus $M$ defines the robustness of the code, ordinary residue codes only
guarantee the detection of a single bit flip, because a single bit flip on the
data and on the residue is enough to construct a new, valid code word (e.g., the
Hamming distance between the $0_c$ and $1_c$ is two, where both values denote
a residue encoding with an arbitrary modulus $M$).

In order to overcome this limitation and to scale the robustness of the code,
the redundancy part can be increased by using more than one
residue~\cite{DBLP:journals/tc/Rao70,DBLP:journals/tit/RaoG71}, yielding a
multi-residue code. The modulus $M$ is now defined by $M = \left\{m_0, \dotsc,
m_n \right\}$, where $n$ is the number of residues.

Finding the set of moduli is not easy. Although finding good moduli requires
exhaustive search~\cite{DBLP:conf/wisa/MedwedS09}, the moduli selection for
multi-residue codes can be done faster than for AN-codes. Residue codes, in
general, are arithmetic codes and therefore also support certain arithmetic
operations. Here, the operation is performed on the functional part and on the
residues independently. \autoref{eq:res_add} shows how an addition works for two
multi-residue encoded values. The addition is performed on the functional value
and on the residues independently. On the residues, the addition is performed
followed by a modular reduction using on the moduli $m_i$ for the
i\textsuperscript{th} residue.

\begin{equation}
	z_c = x_c + y_c = \left(x + y , \forall i: m_i | \left(r_{i,x} + r_{i,y}\right) \right)
	\label{eq:res_add}
\end{equation}

Similar to the addition, residue codes also support subtractions and
multiplications. However, in this work, we only use additions and subtractions.

\subsection{ARM Pointer Authentication}

Protecting pointers against tampering is not only relevant in the context of
fault attacks but is also used to counteract software attacks. For example, ARM
added a feature called pointer
authentication~(PAC)~\cite{arm2017pointerauthentication} to the \textit{ARM
v8.3} instruction set with the goal of protecting special pointers. Several new
instructions were added to the architecture that permits to cryptographically
authenticate special pointer values in registers, like the return address in the
link register, using a message authentication code~(MAC). PAC tags have a size
between 3 and 31 bits, depending on the processor configuration, and are, as in
our work, directly embedded into the protected pointer.

Note, however, that even though the general approach is similar to our work, the
provided capabilities and the resulting protection is vastly different. PAC aims
to only protect special pointers against software attacks. In PAC, authenticated
pointers cannot be be protected during pointer arithmetic since there is no
homomorphism for the MAC. Furthermore, PAC only aims to protect the pointer. The
memory access, which uses an authenticated pointer is completely unprotected
and there are no protection mechanisms to ensure that the accessed memory
actually originates from the correct address.

\section{Pointer Protection with Residue Codes}
\label{sec:pointer}

Manipulation of a memory access is possible by attacking two different parts of
the access. The first one is the pointer itself, which is used to perform the
memory access. This section details how we use multi-residue codes to protect
every data pointer inside a program against fault attacks. Furthermore, we present
how to integrate the multi-residue code into our pointer representation and
elaborate on the additionally needed hardware support.

\subsection{Overview}

Pointers are ubiquitous. Every memory access, e.g., accessing a variable on the
stack, uses a pointer to address the memory. However, when considering fault
attacks, pointers may be manipulated to point to a different memory location.

To counteract this threat, we encode all pointers to a redundant representation,
where faults are detectable. As presented previously, there are two classes of
suitable codes: \textit{separable} and \textit{non-separable} codes, which have
similar properties in terms of error detection capabilities and support for
computation. However, a separable code has advantages to protect a pointer.
Namely, it supports direct access to the functional value and can therefore
immediately be used to address memory. On the other hand, using a non-separable
code to protect the pointer requires to perform a potentially expensive decoding
operation before the actual address is available. AN-codes, as an example for
non-separable codes, require a costly integer division during the decoding
operation. Hence, this division would be required for every memory access.

We encode pointers using a \textit{separable} multi-residue code with a scalable
number of moduli. Here, an encoded pointer $p_c$ is denoted as a tuple $(p,
r_p)$, where $p$ is the original value of the pointer and $r_p$ denotes the
redundancy part comprising the residues of $p$ given a moduli set $M$. Using a
multi-residue code to protect the pointer gives two advantages. On the one hand,
the strength of the code, i.e., the number of bit flips which are detectable, is
scalable with the number of residues. On the other hand, residue codes are
arithmetic codes and therefore also support arithmetic instructions, like
addition and subtraction, natively. This allows us to perform pointer
arithmetic, for example, the stack pointer manipulation in function prologues
and epilogues, directly inside the encoded domain without decoding the pointer.

\subsection{Pointer Layout and Residue-Code Selection}

Adding separable redundancy to data implies that the additional information
needs to be stored somewhere in order to provide a value. In the context of
protecting a processor register, various possibilities exist to provide this
storage.

For example, an additional parallel register file can be added to the processor,
which only holds the redundancy part and gets updated in lockstep with the
actual values~\cite{DBLP:conf/date/MedwedM11}. However, this approach is quite
costly for our use case considering that only a small number of registers
typically hold pointers at a certain point in time. Alternatively, pairs of
regular registers can be used to store the data and its redundancy.
Unfortunately, doing so increases the register pressure and lowers the overall
performance. Moreover, without adding costly access ports to the register file,
multiple instructions have to be performed on every pointer operation, even for
simple ones like an increment. Finally, at least for modern RISC instruction set
architectures~(ISA)~\cite{Waterman14therisc-v}, adding additional operands into
the instruction encoding is difficult without increasing the instruction size.

In this work, we therefore went with a different approach and stored the
redundancy information directly into the upper bits of the pointer. Similar to
PAC, i.e., ARM's pointer authentication feature, this approach introduces zero
overhead in terms of storage for the redundancy at the cost of some bits of
address space. Additionally, this dense representation of an encoded pointer
allows us to add new combined residue arithmetic instructions, which operate on
the functional value and on the residues in parallel, rather than requiring
separate instructions to handle both. By storing the redundant pointer in one
register, we can, therefore, use the same instruction format as regular
instructions and do not require extensive modifications of the ISA or hardware
to maintain performance.

Considering that the directly accessible address space is limited, embedding the
residues into the pointer works best for modern 64-bit architectures. Therefore,
the following design considerations as well as our prototype, that is presented
in \autoref{sec:architecture}, is built upon such an architecture. The overall
concept can still be applied to 32-bit architectures with reduced error
detection capabilities or via a different storage option.

\paragraph{Parameter Selection.}

When selecting the parameters of an error detecting code, it is always a
trade-off between error detection capabilities and the overhead introduced by
the code. However, since the functional value including the redundancy is stored
in a single register, also the remaining address space has to be considered. For
our prototype, we focus on a 64-bit architecture and partition our pointers into
24-bit redundancy and a 40-bit functional value. The resulting pointers can
still address one terabyte of memory, which is sufficient for most applications.

As a concrete code, we instantiate a multi-residue code with the moduli set $M =
\left\{5, 7, 17, 31, 127\right\}$, which is an extension to the one presented
in~\cite{DBLP:conf/date/MedwedM11}. This moduli set yields a code with a Hamming
distance of $D=5$ and is capable of detecting up to four bit flips in the
encoded 64-bit pointer value. Storing the residues for these moduli requires a
total of 23 bits, i.e, $3,3,5,5,7$ bits, respectively. The last remaining bit is
used as a tag bit and specifies if data accessed via the pointer have to perform
data linking/unlinking as presented later in \autoref{sec:mmio}. The resulting
register layout of such an encoded pointer is shown in
\autoref{fig:encoded_ptr_mmio}.

\newcommand{\fakesixtyfourbitsWithRes}[1]{
	\tiny
	\ifnum#1=1234567890
		#1
	\fi
	\ifnum#1=0
		0
	\fi
	\ifnum#1=19
		39
	\fi
	\ifnum#1=20
		40
	\fi
	\ifnum#1=23
		43
	\fi
	\ifnum#1=26
		46
	\fi
	\ifnum#1=30
		51
	\fi
	\ifnum#1=34
		55
	\fi
	\ifnum#1=39
		63
	\fi
}

\begin{figure}
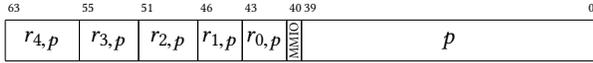

	\centering
	\begin{bytefield}[
		bitwidth=\widthof{\rotatebox{90}{\tiny MMIO}~},
		boxformatting={\centering},
		bitformatting=\fakesixtyfourbitsWithRes,
		endianness=big]{40}
		\bitheader{39,34,30,26,23,20,19,0} \\
		\bitbox{5}{$r_{4,p}$} &
		\bitbox{4}{$r_{3,p}$} &
		\bitbox{4}{$r_{2,p}$} &
		\bitbox{3}{$r_{1,p}$} &
		\bitbox{3}{$r_{0,p}$} &
		\bitbox{1}{\rotatebox{90}{\tiny MMIO}}  &
		\bitbox{20}{$p$}
	\end{bytefield}
	\caption{Encoded pointer representation. The actual 40-bit pointer value $p$,
           the MMIO tag bit, and 23 bits of redundancy $r_p$ comprise an encoded
           64-bit pointer.}
	\label{fig:encoded_ptr_mmio}
\end{figure}

\subsection{Pointer Operations}
\label{sec:pointer_operations}

Pointers are not only used to perform a memory access but also are used to
perform pointer arithmetic. To maintain good performance, it is therefore vital
that the encoded pointers support these computations as efficiently as possible.
Notably, as the term pointer arithmetic already hints, arithmetic computations,
like addition and subtraction, are the most common operations that are performed
on pointers. For example, accessing larger sequential memory chunks via a
pointer involves a large number of additions between the pointer and the access
stride in a loop. Similarly, next to every function call, the respective stack
frame size is added and subtracted to/from the stack pointer in the function's
prologue and epilogue. Precisely these types of operations are natively
supported by the used multi-residue code and can therefore be performed in the
encoded domain.

On the other hand, more work is required for operations that are not directly
supported by the multi-residue code. The simplest approach is probably to
perform the operation on the plain functional value only and restore the
encoding afterward. To ensure the correctness of the computation, then
additional measures like replication have to be used. Alternatively, such
operations can be performed by first converting the pointer to a different code,
in which the computations are straightforward, followed by converting the
differently encoded result back into multi-residue representation. Still, such
operations comprise only a very small number of pointer operations compared to
arithmetic operations.


\paragraph{Software vs. Hardware.}

In a multi-residue code, the addition operation is performed on the functional
value and on all its residues. This operation can be executed in hardware or
software. However, performing this operation in software is challenging, as it
involves a modulo reduction for each residue.

Looking only at a single modulo, there exist several options for implementing
the reduction in software:
First, a normal modulo instruction from the ISA can be used. Although such an
instruction does not have much code overhead, a modulo operation involves a
costly integer division which usually takes multiple clock cycles to finish.
Second, instead of a modulo operation, a conditional subtraction can be used for
the modular reduction. Third, there are optimized modulo algorithms
available~\cite{jones2001modulus}, but their overhead is still large. A single
modular addition with an optimized reduction with the modulus five takes at
least 18 instructions on our \riscv target architecture.

Considering that the runtime of these solutions additionally has to be
multiplied with the number of used residues makes a software solution even less
attractive. Furthermore, even if the performance penalty is acceptable,
additional registers have to be reserved for implementing the reduction
functionality. Summarizing, a software-based approach to perform residue
operations, while feasible, is not very practical. Therefore, hardware based
approaches to implement the residue operations have been investigated.

In particular, in our prototype, we add new instructions that permit to perform
addition and subtraction of multi-residue encoded pointers.
\autoref{sec:hw_instructions} discusses the new instructions in detail with the
focus on the target architecture. Furthermore, an instruction for performing the
expensive encoding operation is added, which computes the modulus for each
residue. For convenience reasons, also a dedicated decoding operation is added
to the ISA.

\section{Evolved Memory Access Protection}
\label{sec:memory}

Apart from faulting the pointer, the second source to manipulate a memory access
is the memory operation itself. If the attacker is able to induce faults on the
address bus, the memory access can be redirected to a different location. In
this section, we present a method to link the data with its respective address,
where addressing errors are transformed into data errors which can subsequently
be detected using a data-protection scheme.

\subsection{Overview}

In order to be able to detect address tampering, a way to uniquely identify
incorrectly accessed memory is needed. A common approach to establish this link
between the data and the address is augmenting the data-protection scheme, which
is anyway needed to protect the data against faults.

For example, ANB-codes embed the identity of the variable, in the form of a
unique residue $B_x$, into a required underlying AN-code based data encoding.
However, this approach has several drawbacks. For example, working on variable
granularity requires concise data-flow information, which is in real-world
applications hard to acquire for arbitrary memory operations, and limits the
applicability of the approach. Furthermore, maintaining these identities during
calculation is quite costly. Finally, the approach is strongly linked with
AN-codes and cannot easily be applied to other data-protection schemes.

Our scheme, takes an entirely different approach to prevent address tampering.
Instead of constructively embedding the address of the data into the
data-protection code, our scheme destructively overlays data that is written to
the memory with the respective memory address. As a result, addressing errors
are transformed into data errors that get detectable as soon as the overlay is
removed again.

In more detail, before data is written from a register to the memory bus by the
processor, the data gets encoded with respect to the target address.
Conceptually, this kind of linking is similar to encrypting the data in an
address dependent way. However, since we do not strive for confidentiality with
our approach, the use of a cryptographically secure cipher is not needed. The
resulting encoded data is then simply stored into memory like in a regular
system.

When data is read back from memory into a processor register, the decoding with
respect to the target address is performed. Considering that the performed
decoding operation is the inverse of the encoding, a genuine data value is
restored only when the read has been performed from the correct address.
Otherwise, an incorrect data value is generated which can be detected via the
used data-protection scheme. Note that the detection of address tampering during
memory writes is possible like this as well. However, the detection is delayed
to the point where the incorrectly written value is read back into the
processor.


\subsection{The Linking Approach}

As already mentioned, the general idea behind our memory access protection
approach is to link the data that is stored in memory with its respective
address. Instead of directly writing a register value $D_{Reg}$ into memory at
the a certain address $p$ (i.e.,~$\mathtt{mem}\left[p\right] = D_{Reg}$), a
little more work has to be performed in our scheme. Namely, as shown in
\autoref{eq:correction}, the linking function $l$ has to be evaluated in order
to determine the value that is actually written to the memory at address $p$.

\begin{equation}
	\mathtt{mem}\left[p\right] = l\left(p, D_{Reg}\right) = l_p\left(D_{Reg}\right)
	\label{eq:correction}
\end{equation}

The purpose of this linking function is to combine the address $p$ with the data
value $D_{Reg}$. However, not every function can be used for this purpose. At
the very least, the following two requirements have to be fulfilled in this
context. First, for each address $p$, the linking function $l_p$ has to be a
permutation. Having this property means that $l_p$ performs a bijective mapping
and that an inverse function $l_p^{-1}$ exists, i.e.,~$\forall p, D_{Reg}
\rightarrow l_p^{-1}\left(l_p\left(D_{Reg}\right)\right) = D_{Reg}$.
Subsequently, memory read operations can be implemented using this inverse
function as shown in \autoref{eq:inv_correction}. As the result, from the
software perspective, encoding data when storing to memory and decoding data
when loading from memory is completely transparent, yields the expected result,
and can be performed for every memory access.

\begin{equation}
	D_{Reg} = l^{-1}\left(p, \mathtt{mem}\left[p\right]\right) = l_p^{-1}\left(\mathtt{mem}\left[p\right]\right)
	\label{eq:inv_correction}
\end{equation}

Second, to ensure that addressing faults are detectable, data encoded under one
address should yield a modified value when being decoded under a different
address, i.e.,~$\forall p, p', D_{Reg}: p \neq p' \rightarrow
l_{p'}^{-1}\left(l_p\left(D_{Reg}\right)\right) \neq D_{Reg}$. Note,
furthermore, that the modified value should not be a valid code word in terms of
the used data-protection code.

\paragraph{Function Selection.}

Various functions, like for example cryptographic ciphers, fulfill these
requirements and are therefore suitable to link the data and the address
information as required by the memory access protection scheme. However, given
that we aim for a low-overhead design, less resource demanding functions have
been investigated.

Interestingly, already a simply xor operation, as shown in
\autoref{eq:f_xor_enc} and \autoref{eq:f_xor_dec}, is sufficient as the linking
function for our use case. In more detail, in our scheme, addresses are encoded
using arithmetic multi-residue codes and the data encoding can be selected
arbitrarily. On the one hand, when the same multi-residue code is also used for
the data, e.g., an encoded pointer is written to memory, using the xor operation
is good choice given that that multi-residue codes are not closed under the xor
operation. Subsequently, it is also unlikely that combining multiple valid code
words yields a valid result and therefore facilitates error detection. On the
other hand, even when a data protection code which is closed under the xor
operation is used, still similar error detection capabilities are expected.
After all, combining code words from different codes is highly unlikely to yield
sensible results.

\begin{align}
	\mathtt{mem}\left[p\right] &= p \oplus D_{Reg} \label{eq:f_xor_enc} \\
	D_{Reg} &= p \oplus \mathtt{mem}\left[p\right] \label{eq:f_xor_dec}
\end{align}

\paragraph{Linking Granularity.}

Theoretically, the previously described linking approach can be applied with
arbitrarily granularity. Therefore, applying the technique on the
processor's native word size, e.g., 64-bit in our prototype, may appear
natural. However, performing xor-based linking on such a coarse granularity does
not yield the desired amount of diffusion. Namely, bytes that are close to each
other, i.e., with a stride of 8 bytes when operating on 64-bit, are highly
likely to have the same address pad. Furthermore, in many real-world
applications, also misaligned data accesses with arbitrary size have to be
supported efficiently. Situations like this, for example, commonly arise when
arbitrarily aligned data is copied via the \textit{memcpy} function.

Therefore, to fix the problem of the low diffusion and the arbitrarily aligned
data accesses, we perform the linking of data and address with byte-wise
granularity. Each byte, even when it is part of a larger memory transfer, is
independently linked with its respective address. Hereby, each individual
byte-address pointer is still multi-residue encoded to provide the desired
diffusion during linking. Furthermore, the actual linking is again performed via
an xor similar to \autoref{eq:f_xor_enc}. However, considering that the data and
its address have different bit sizes, an additional compression is applied on
the address before linking. Namely, each 64-bit address $p = [p_0, p_1, \ldots,
p_7]$ gets reduced to an one byte value $p'$ by xor-ing the individual address
bytes as shown in \autoref{eq:ptr_reduce}.

\begin{equation}
	p' = \bigoplus_{i=0}^7 p_i \label{eq:ptr_reduce}
\end{equation}

\begin{figure}[t]
	\centering
	\includegraphics[width=1\columnwidth]{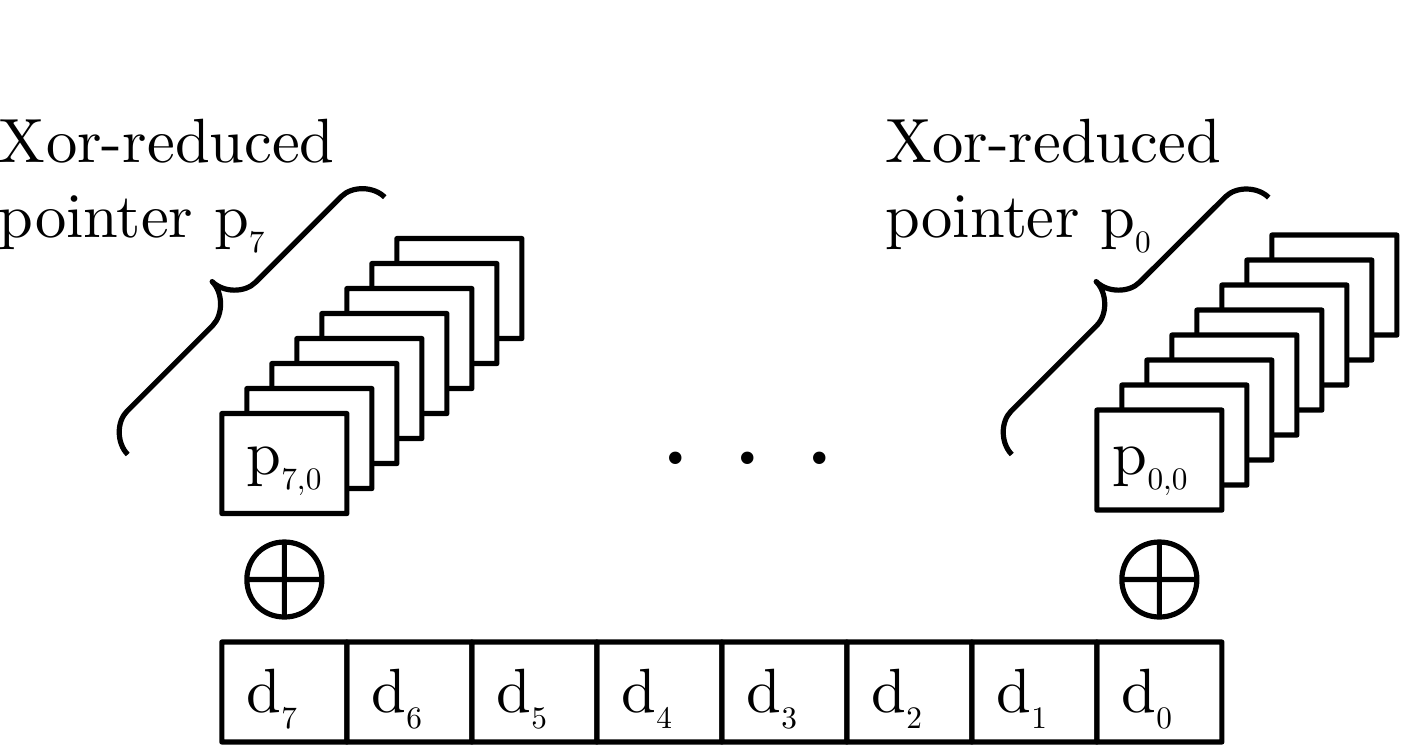}
	\caption{Byte-wise data linking of a 64-bit word. Each byte gets xored with
	its respective xor-reduced encoded address.}
	\label{fig:data_corr_ptr_reduced}
\end{figure}

Applying this approach to a full 64-bit word is visualized in
\autoref{fig:data_corr_ptr_reduced}. Considering the number of needed
multi-residue operations for such a word-sized access, using this linking scheme
effectively requires hardware support. In this work, we therefore integrated the
needed transformations directly into special load and store instructions.
From the software perspective, encoding data when storing to memory and decoding
data when loading from memory is completely transparent and can be performed for
next to every memory access.

\subsection{Memory-Mapped I/O}
\label{sec:mmio}

Memory-mapped I/O~(MMIO) is a common communication interface in embedded
processors to access peripherals. In MMIO, the peripheral registers are mapped
into the standard memory layout of the processor. This allows the CPU to use
ordinary load and store instructions to access the peripheral.

However, in order to protect the memory access, our architecture uses redundant
pointers and links them with the data before executing the memory access. Since
a standard memory-mapped peripheral is not aware of this data linking, wrong
data would be written to the device. Therefore, we cannot apply data linking
when accessing a memory-mapped peripheral. However, we still can use an encoded
pointer to access the memory-mapped peripheral as this does not influence the
data. In order to use an encoded pointer but not perform the data linking, we
would need special instructions for load and store for this purpose. We avoid
this overhead by encoding this information directly into the encoded pointer.
The load and store instructions detect this and do not perform the data linking
accordingly.

As shown in~\autoref{sec:pointer}, we redundantly encode the pointer using a
multi-residue code. In \autoref{fig:encoded_ptr_mmio}, we show the pointer
layout where the 41\textsuperscript{st} MMIO-bit indicates whether the pointer
is for an MMIO access without data linking. The residues, which form the
redundancy of the pointer, are computed over the 40-bit functional pointer value
and the MMIO-bit to protect both against tampering.

\section{Architecture}
\label{sec:architecture}

The concept of protected pointers and linked memory accesses is integrated in a
prototype implementation based on a 64-bit \riscv architecture. In this section
we first discuss the new instructions, show how we integrated them into the
architecture, and finally show a compiler prototype to automatically protect all
memory accesses in a program.

\subsection{New Instructions}
\label{sec:hw_instructions}

As previously described, it requires hardware support to efficiently perform the
residue arithmetic such that the performance penalty is acceptable. In this
work, we extend the instruction set of the processor with instructions that
operate in the encoded residue domain. In particular, the following custom
instructions are added to the instruction set.

\paragraph{\textit{renc}, \textit{rdec}.}

To efficiently encode a value into the multi-residue domain, a dedicated
encoding instruction (\textit{renc}) is added. The encoding operation computes
the residues over the 41-bit functional value of the pointer, which also
includes the \textit{MMIO}-bit in the protection domain. As a second
instruction, we add support to decode a multi-residue encoded register
(\textit{rdec}). Both instructions are idempotent, meaning they can repeatedly
be executed (encoding an already encoded value does not change the encoding).

\paragraph{\textit{radd}, \textit{raddi}, \textit{rsub}.}

To support pointer arithmetic on encoded pointers, hardware support for the most
commonly used operations is added. Concretely, we support adding two
multi-residue encoded register values (\textit{radd}), adding a multi-residue
encoded value to an immediate value (\textit{raddi}), and subtracting
multi-residue encoded values (\textit{rsub}). The immediate value in the
\textit{raddi} instruction is not yet multi-residue encoded. However, these
values are part of the instruction encoding and are already protected via the
CFI code protection scheme. Note that, before the immediate can be used in a
residue operation, it gets encoded as part of the instruction execution.

\paragraph{\textit{rlxck}, \textit{rsxck}.}

Since we now use encoded pointers and require data linking/unlinking, dedicated
memory instructions are added to the ISA. Therefore, a family of new load
(\textit{rlxck}) and store (\textit{rsxck}) instructions is added. Herby, the
\textit{x} denotes the access granularity of the memory operation. Concretely,
we support byte (\textit{b}), half-word (\textit{h}), word (\textit{w}), and
double word (\textit{d}) accesses with and without sign extension, which
corresponds to the original memory access instructions in the \riscv 64-bit ISA.
The new instructions have the same operand interface as the original load and
store instructions of \riscv. However, they now take an encoded pointer for
addressing the memory. The memory instructions also contain a plain immediate
value to add an offset to the pointer, which is protected by the CFI code
protection. Furthermore, these instructions perform the data linking and
unlinking on a byte-wise granularity. However, if the
40\textsuperscript{th}-bit, the \textit{MMIO} bit, is set to one, no data
linking and unlinking is performed, which allows us to use a protected pointer
when accessing a memory location, which does not support data linking, e.g., a
memory-mapped peripheral.

Since every memory access is replaced with its protected counterpart, the
protection mechanism could already be implemented in the original load and store
instructions of the processor. However, for the sake of still supporting the
original \riscv instructions, they are left unmodified, and new instructions are
added separately.

\subsection{Hardware}

\begin{figure}[t]
	\centering
	\includegraphics[width=1\columnwidth]{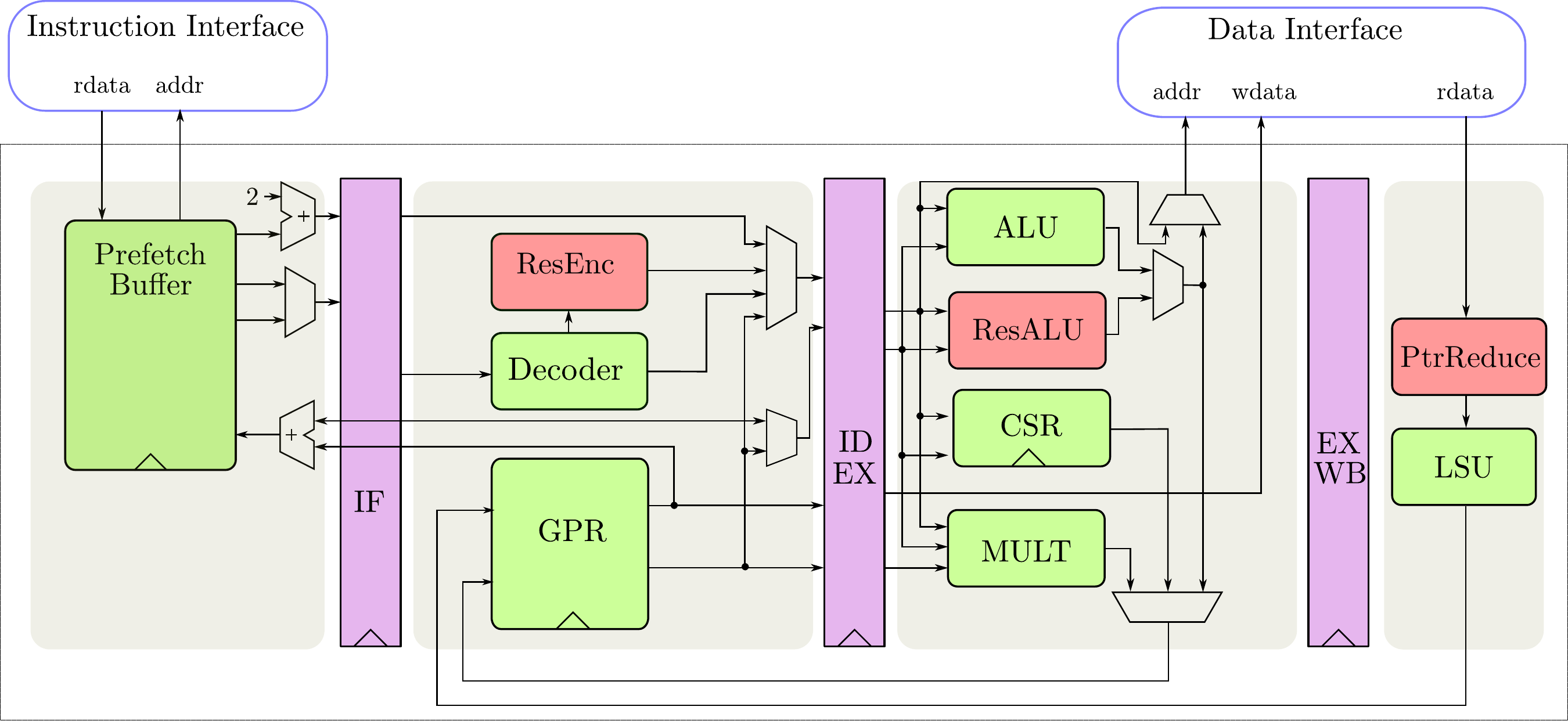}
	\caption{Modified processor pipeline. The instruction decode stage is extended
					 with a 12-bit residue encoder, the execution stage with a residue
					 ALU, and the write-back stage with a pointer-reduction data-linking
					 unit.}
	\label{fig:residue_pipeline}
\end{figure}

The instruction set is only one part of our protected architecture. We also
implemented the modified instruction set in hardware. As foundation, the
open-source 32-bit \riscv core \textit{\riscy}~\cite{pulp2018riscy} is used.
This core is extended to a 64-bit processor meaning that the register file,
datapath, and load-and-store unit are modified and all necessary instructions
are added to be compliant with the \riscv RV64IM instruction set.
Furthermore, we added the new instructions to deal with multi-residue encoded
pointers, as defined in \autoref{sec:hw_instructions}.
\autoref{fig:residue_pipeline} shows the modified processor pipeline, which
includes a dedicated arithmetic logical unit~(ALU) for residue operations.
Furthermore, immediate values, which are part of the instruction, get encoded
during the instruction decode stage of the processor pipeline. The
load-and-store unit is extended to support data linking and unlinking to protect
all memory accesses.

\begin{figure}[t]
	\centering
	\includegraphics[width=0.8\columnwidth]{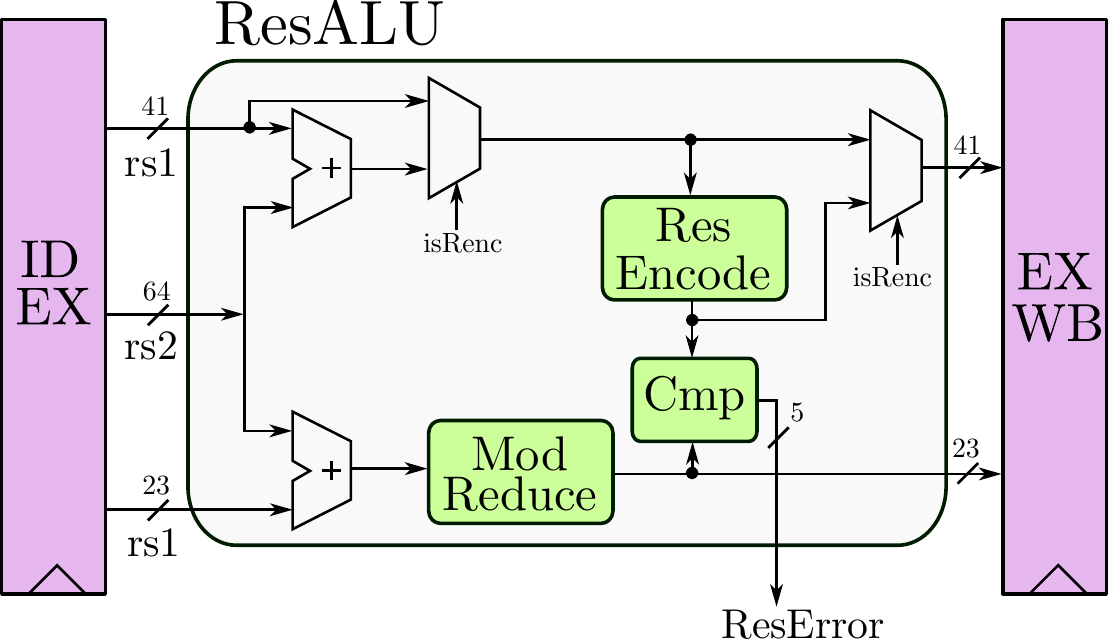}
	\caption{Residue ALU with a 41-bit adder and a shared residue encoder. The
					 addition result is automatically checked after the operation by
					 re-encoding the result and comparing it with the computed residues
					 and generating a redundant error signal.}
	\label{fig:residue_alu}
\end{figure}

The new residue ALU is shown in detail in \autoref{fig:residue_alu}. The ALU
supports encoding and decoding of values to/from the multi-residue domain as
well as adding and subtracting two encoded values. The design of the ALU is
optimized to require only one residue adder and one encoder in the execution
stage of the processor. Decoding is for free since it only requires rewiring,
where the upper bits are set to zero. After performing an addition, the
functional value of the adder result is re-encoded and compared with the
independently computed residues in order to perform error-checking after each
residue instruction. If the computed residues and the newly re-encoded residues
mismatch, a redundant error signal is generated to force the processor into a
safe state. Since this adder is also used for computing the final pointer
address during a memory access (the encoded immediate value is added to the
encoded base pointer), every pointer is also checked before performing a memory
access. With frequent checks for every result, we minimize the probability that
error masking occurs and errors are not detectable anymore.

Currently, the residue encoder uses special algorithms
from~\cite{DBLP:journals/vlsisp/PerssonB09} to encode data. However, the residue
adder is implemented without any further optimizations. By using optimized
arithmetic operations, e.g., the one from~\cite{DBLP:conf/arith/Zimmermann99},
the hardware overhead can be further reduced.

\subsection{Software}

To make the countermeasure practical and protect every memory access in the
program, the new instructions and the protection mechanism also need to be
integrated into the compiler. In the following, we integrate our countermeasure
to the LLVM-based C compiler~\cite{DBLP:conf/cgo/LattnerA04}.

An LLVM-based compiler is partitioned into three parts, the front end, the
middle end, and the back end. While the middle end optimizes
target-independently on an intermediate code representation, the back end
transforms the universal intermediate representation to a target-dependent code.
To protect every memory access in the program, the countermeasure needs to be
inserted in the back end stage of the compiler. Any earlier transformation can
potentially miss a memory access leaving some accesses possibly unprotected
(e.g., the stack is created in the target-dependent part of the back end). Even
in the back end, the protection happens right before the final instruction
scheduling.

LLVM's back end uses a directed acyclic graph~(DAG) representation, the
\textit{Selection DAG}, for the code generation. The intermediate representation
is transformed in a series of steps to finally emit the machine code. However,
the back end has no information about pointers and addresses. Therefore, this
information is created and propagated manually on the Selection DAG. Dedicated
pointer nodes are added to the Selection DAG where pointers are created, e.g.,
when creating a \textit{FrameIndex} node used for a local stack memory access.
This information is then propagated on the Selection DAG and all dependent
operations are replaced with their corresponding residue counterpart. If we
obtain an instruction, which is not supported by the residue code, the pointer
is decoded, the operation is performed in the unencoded domain, and then, the
pointer is re-encoded. However, this sequence of instructions is not used in the
majority of the transformations. Finally, protected load and store instructions
are emitted, which use an encoded pointer for addressing the memory.

If the program uses a constant address, e.g., the address of a global variable,
this information needs to be encoded to the multi-residue domain. However, the
compiler does not have this information yet. Therefore, it creates a relocation
such that the linker can fill in the correct address information. Since this
information requires multi-residue encoding, the linker is also modified. In our
work, we use a custom \riscv back end of LLVM's \textit{lld} linker. In addition
to resolving regular relocations, our linker also applies multi-residual
encoding to pointers in the binary. This includes pointers synthesized in code
as well as pointers stored in the memory, which additionally get linked with
address information. Similar to that, data stored in the read-only section of
the binary is also linked with its address. As soon as these values are loaded
into a register, the unlinking operation is performed and the correct value is
restored.

\section{Evaluation}
\label{sec:evaluation}

In order to make a countermeasure usable in practice, the overhead must be
reasonable. In this section, we first show the introduced hardware overhead and
then evaluate different benchmark applications on the target architecture.
Finally, we analyze the software overhead, discuss the overhead sources, and
describe future optimization possibilities.

To quantify the hardware overhead, we synthesize the hardware architecture for a
Xilinx Artix-7 series FPGA. By adding the new instructions, a dedicated ALU for
multi-residue operations, and a modified load-and-store unit, the required
number of lookup-tables~(LUTs) increases by less than 5\,\%, and the number of
flip-flops increases by less than 1\,\%. However, this prototype design is
implemented without optimizations leaving space to further improve the design.

The custom LLVM toolchain based on LLVM 6.0 is used to compile different
benchmark applications for the \riscv-based target architecture. The benchmarks
were taken from the \textit{PULPino} repository~\cite{pulp2018pulpino}, which
were used to originally evaluate the performance of the \riscy core. Simulation
is performed using a cycle accurate HDL simulation of the target processor. As
baseline, we simulate the benchmark applications solely with enabled CFI
protection~\cite{DBLP:journals/corr/abs-1802-06691} but without an
application-specific data protection scheme. On top of that baseline, we
determine the exclusive overhead of our countermeasure in terms of code size and
runtime.

As shown in \autoref{tab:overhead}, on overage, the code overhead is 10\,\% and
the runtime overhead is less than 7\,\%. This is a comparable better performance
to ANB-codes, which have an average runtime overhead of 90\,\% compared to
AN-codes solely to provide memory access protection. Instead, our countermeasure
has a considerable lower overhead, making it attractive for many real-world
applications.

\begin{table}[t]
	\centering
	\caption{Code and runtime overhead for different benchmark programs from an HDL simulation.}
  \label{tab:overhead}
	\begin{tabular}{c|cc|cc}
		\toprule
		\multirow{3}{*}{Benchmark} & \multicolumn{2}{c}{Code Overhead} & \multicolumn{2}{c}{Runtime Overhead} \\
															 & Baseline  & Overhead & Baseline   & Overhead \\
															 & [kB]      & [\%] & [kCycles]      & [\%] \\
		\midrule
    fir & 4.26 & 8.54 & 39.22 & 6.35 \\
    fft & 6.52 & 6.57 & 58.01 & 4.65 \\
    keccak & 4.79 & 10.11 & 255.55 & 11.31 \\
    ipm & 4.84 & 12.81 & 10.80 & 3.94 \\
    aes\_cbc & 7.25 & 8.77 & 60.91 & 9.10 \\
    conv2d & 3.26 & 13.12 & 5.92 & 2.70 \\
    \midrule
    Average & & 9.99 & & 6.34 \\
    \bottomrule
  \end{tabular}
\end{table}

\subsection{Future Work}

The overhead numbers are already competitive for practical usage. Still, some
improvements regarding code size or performance have not been performed yet.

For example, pointer comparisons in the encoded domain are currently only
implemented for \textit{equal} and \textit{not equal}. Although seldomly used,
comparisons with other predicates are still performed on the functional value.
Similarly, in are rare cases, when pointer arithmetic uses unsupported logical
operations, the operations is performed on the functional value only. Adding
support for these operations further reduces the overhead and slightly increases
the protection domain.

Furthermore, our current toolchain has not been highly optimized for our
prototype architecture yet. We expect that, with a more optimized compiler, even
better results can be achieved in the future.

\section{Conclusion}
\label{sec:conclusion}

Memory accesses are frequently used operations, and many different security
policies, as well as safety mechanisms, rely on their correct execution.
However, when dealing with faults, a correctness of a memory access cannot be
guaranteed. While there are dedicated methods to protect the control-flow of a
program and to protect the data in memory and registers, there is no efficient
protection mechanism to protect the memory access against address tampering.

In this work, we closed this gap and presented a new mechanism to protect memory
accesses inside a program. The countermeasure is employed in two steps. First,
all pointers including pointer arithmetic are protected by employing a
multi-residue code. The redundancy is hereby directly stored inside the unused
upper bits of the pointer, which does not add any memory overhead. The second
step links the redundant pointer with the data. Subsequently, addressing errors
manifest as data errors and get detectable as soon the data is loaded into the
register. This linking approach is universally applicable and can be used on top
of any data protection scheme.

To demonstrate the practicability of our countermeasure, we integrated the
concept of protected memory accesses into a RISC-V processor. We extended the
instruction set to deal with multi-residue encoded pointers and added new memory
operations which perform the linking and unlinking step. Furthermore, we
extended the LLVM compiler to automatically transform all pointers of a program
to the encoded domain. Our evaluation showed an average code overhead of 10\,\%
and an average runtime overhead of less than 7\,\%, which makes this
countermeasure practical for real-life applications.

\ifauthor
\section{Acknowledgment}
\label{sec:Acknowledgement}

This project has received funding from the European Research Council (ERC) under
the European Union's Horizon 2020 research and innovation programme (grant
agreement No 681402) and by the Austrian Research Promotion Agency (FFG) via the
competence center Know-Center (grant number 844595), which is funded in
the context of COMET - Competence Centers for Excellent Technologies by BMVIT,
BMWFW, and Styria.

\fi

\balance
\bibliographystyle{ACM-Reference-Format}
\bibliography{bib/bibliography.bib}

\end{document}